\documentstyle[12pt]{article}
\advance \textheight by \topskip
\makeatletter
\def\eqnarray{\stepcounter{equation}\let\@currentlabel=\theequation
\global\@eqnswtrue
\global\@eqcnt\z@\tabskip\@centering\let\\=\@eqncr
$$\halign to \displaywidth\bgroup\@eqnsel\hskip\@centering
  $\displaystyle\tabskip\z@{##}$&\global\@eqcnt\@ne 
  \hfil$\displaystyle{\hbox{}##\hbox{}}$\hfil
  &\global\@eqcnt\tw@ $\displaystyle\tabskip\z@
  {##}$\hfil\tabskip\@centering&\llap{##}\tabskip\z@\cr}
\@addtoreset{equation}{section}
  \def\theequation{\thesection.\arabic{equation}}
\makeatother

\mathchardef\by="0202

\begin{document}

\title{Universality of the $R$-deformed Heisenberg 
algebra\footnote{\it Invited talk at 
the International Seminar ``Supersymmetry and Quantum Field Theory"
dedicated to the memory of
D. V. Volkov, Kharkov (Ukraine), January 5-7, 1997.}} 
\author{Mikhail S. Plyushchay\\
\smallskip\\
{\small \it Institute for High Energy Physics,
Protvino, Moscow Region, 142284 Russia}\\
{\small \it Departamento de Fisica -- ICE,
Universidade Federal de Juiz de Fora}\\
{\small \it 36036-330 Juiz de Fora, MG Brazil}\\
{\small \it E-mail: plyushchay@mx.ihep.su}}

\date{}

\maketitle

\begin{abstract}
We show that deformed Heisenberg algebra with reflection
emerging in parabosonic constructions is also related to
parafermions. This universality is discussed in different
algebraic aspects and is employed for the description of
spin-$j$ fields, anyons and supersymmetry in 2+1 dimensions.
\end{abstract}

\section{Introduction}

The $R$-deformed Heisenberg algebra (RDHA)
is given by the generators $a^-$, $a^+$, $R$, and $1$
satisfying the (anti)commutation relations
\begin{equation}
[a^-,a^+]=1+\nu R,\quad
\{a^\pm,R\}=0,
\quad
R^2=1,\quad
[1,a^\pm]=[1,R]=0,
\label{1}
\end{equation}
where $\nu\in {\bf R}$ is a deformation parameter and $R$ is 
a reflection operator.
It emerges in the context of quantization schemes
generalizing bosonic commutation relations as follows.
Let us consider a quantum mechanical system 
having the bosonic-like Hamiltonian, $H=\frac{1}{2}\{a^+,a^-\}$,
and bosonic-like equations of motion,
\begin{equation}
\frac{1}{2}[\{a^-,a^+\},a^\pm]=\pm a^\pm,
\label{2}
\end{equation}
and put the question: what is the most general form
of commutation relations for the operators
$a^+$ and $a^-$ which would lead to eqs. (\ref{2})?
The answer is given by the $R$-deformed 
Heisenberg algebra (\ref{1}) \cite{ok,np1}.
In this way, in fact, the deformed Heisenberg
algebra (\ref{1}) was introduced by Wigner \cite{wig}.

Trilinear commutation relations (\ref{2}) characterize the
{\it parabosonic} system of order $p=1,2,\ldots$ in the case
when $\nu=p-1=0,1,\ldots$ \cite{ok,mac2}.
Generalization of these trilinear commutation relations
to the systems with many degrees of freedom as well as
construction of their fermionic modification
lead Green and Volkov to the discovery  of parafields
and parastatistics \cite{green,volkov,gg}.
Recently the algebra  (\ref{1}) was rediscovered in the context
of integrable systems \cite{poly} 
where it was used for solving quantum mechanical Calogero model
\cite{poly,calog} (see also ref. \cite{yang}). It was also employed 
for bosonization of supersymmetric quantum mechanics 
\cite{mac1}--\cite{ann} and for describing anyons 
\cite{ann,plb0} within the framework of
the group-theoretical approach \cite{plb1}--\cite{volsor}. 

The existence of infinite-dimensional unitary representations of
algebra (\ref{1}) on the half-line $\nu>-1$ and the
relationship between trilinear commutation relations and
$R$-deformed Heisenberg algebra mean that the
latter can be considered as the algebra supplying us with some
generalization of parabosons for the case of non-integer
statistical parameter $p=\nu+1>0$ \cite{mac2}.
But it turns out that the algebra (\ref{1}) obtained originally by
generalizing {\it bosonic} commutation relations, has also
finite-dimensional representations of the deformed
(para){\it fermionic} nature \cite{np1,mpla2}.  
Thus, the $R$-deformed
Heisenberg algebra  reveals some properties of universality
which are the subject of the present talk based on 
recent papers \cite{np1,mpla2,mpla?}.

\section{Aspects of universality}

Infinite-dimensional unitary representations of the algebra
(\ref{1}) taking place for  
$
\nu>-1
$
can be realized on the Fock space with complete orthonormal
basis  of states $|n\rangle=C_n(a^+)^n|0\rangle$,
$n=0,1,\ldots,$  $a^-|0\rangle=0$,
$\langle 0|0\rangle=1$,
$R|0\rangle=|0\rangle$,
where
$C_n=([n]_\nu!)^{-1/2}$,  
$[n]_\nu !=\prod_{l=1}^n [l]_\nu$, $[l]_\nu=l+\frac{1}{2}(1-(-1)^l)\nu$.
The reflection operator is realized as
$R=(-1)^N$, $N=\frac{1}{2}\{a^+,a^-\}-\frac{1}{2}(\nu+1)$,
$N|n\rangle=n|n\rangle$, and 
introduces $Z_2$-grading structure in the space of states,
$
R|k\rangle_\pm=\pm|k\rangle_\pm,
$
$
|k\rangle_+=|2k\rangle,
$
$
|k\rangle_-=|2k+1\rangle,
$
$
k=0,1,\ldots.
$
The even, `$+$', and odd,
`$-$', subspaces are separated by the projectors 
$
\Pi_\pm=\frac{1}{2}(1\pm R).
$

Due to the commutation relation
\begin{equation}
[a^-,(a^+)^n]=\left(n+\frac{1}{2}(1-(-1)^n)\nu R
\right)(a^+)^{n-1},
\label{3}
\end{equation}
at special values of the deformation parameter, $\nu=-(2p+1)$,
$p=1,2,\ldots$, the relation $\langle\langle m|n\rangle\rangle=0$,
$|n\rangle\rangle\equiv (a^+)^n|0\rangle$,
holds for $n\geq 2p+1$ and arbitrary $m$.
Therefore, there are $(2p+1)$-dimensional irreducible
representations of the algebra (\ref{1}) 
with $\nu=-(2p+1)$, in which the relations $(a^+)^{2p+1}=
(a^-)^{2p+1}=0$ are valid. The latter relations are a characteristic
property of parafermions of order $2p$, and
we arrive at the nilpotent algebra
\begin{equation}
[a^-,a^+]=1-(2p+1)R,
\quad
(a^\pm)^{2p+1}=0,\quad
\{a^\pm,R\}=0,\quad
R^2=1,\quad
p=1,2,\ldots.
\label{4}
\end{equation}
Operator $a^+$ can be interpreted here as a paragrassmann
variable $\theta$, $\theta^{2p+1}=0$, whereas 
operator $a^-$ can be considered as corresponding
differentiation operator $\partial$ defined 
by relation (\ref{3}). Therefore, the algebra (\ref{1})
at $\nu=-(2p+1)$ can be considered as a paragrassmann
algebra of order $2p$ 
\cite{fili} with a special differentiation 
operator. It was  called in ref. 
\cite{mpla2} the $R$-paragrassmann algebra.

The finite-dimensional representations 
can be realized as matrix representations
with diagonal operator $R=diag(+1,-1,+1,\ldots,-1,+1)$, 
and with operators $a^\pm$ realized as
$
(a^+)_{ij}=A_j\delta_{i-1,j},\quad
(a^-)_{ij}=B_i\delta_{i+1,j},
$
where
$
A_{2k+1}=-B_{2k+1}=\sqrt{2(p-k)},
$
$
k=0,1,\ldots,p-1,
$
$
A_{2k}=B_{2k}=\sqrt{2k},
$
$
k=1,\ldots,p.
$
Operators $a^+$ and $a^-$ are mutually conjugate,
$(\Psi_1,a^-\Psi_2)^*=(\Psi_2,a^+\Psi_1)$, 
with respect to
the indefinite scalar product 
\begin{equation}
(\Psi_1,\Psi_2)=\bar{\Psi}_{1n}\Psi_{2}^n,
\quad
\bar{\Psi}_n=\Psi^{*k}\hat{\eta}_{kn},
\label{5}
\end{equation}
where $\Psi^n=\langle n|\Psi\rangle$ and
$\hat{\eta}=diag(1, -1, -1, +1, +1,\ldots,
(-1)^{p-1}, (-1)^{p-1}, (-1)^p, (-1)^p)$
is indefinite metric operator.

Finite-dimensional representations of RDHA can also be described 
in terms of hermitian conjugate operators 
$f^+=a^+$, $f^-=a^-R$ and of positive definite
scalar product, $\langle \Psi_1,\Psi_2\rangle=
\Psi^{*n}_{1}\Psi_{2}^n$.
In terms of these, nilpotent algebra (\ref{4}) is 
\begin{equation}
\{f^+,f^-\}=(2p+1)-R,\quad
(f^\pm)^{2p+1}=0,\quad
\{R,f^\pm\}=0,\quad
R^2=1.
\label{6}
\end{equation}
Operators $I_+=f^+$, $I_-=f^-$ generate a nonlinear 
deformation of $su(2)$ algebra of the form
\begin{equation}
[I_+,I_-]=2I_3(-1)^{I_3+p},\quad
[I_3,I_\pm]=\pm I_\pm
\label{7}
\end{equation}
involving the reflection operator $R=(-1)^{I_3+p}$.
For $p=1$ one has the relation $I_3(-1)^{I_3+1}=I_3$,
and in this particular case 
the deformed $su(2)$ algebra turns into the standard
$su(2)$.  
Relations (\ref{7}) can be presented as \cite{np1}
\begin{equation}
[[f^+,f^-],f^\pm]=2I_3(2I_3\mp 1)(-1)^{I_3+p}f^\pm
\label{8}
\end{equation}
with $I_3=C(-1)^{C+p}$, $C=\frac{1}{2}[f^+,f^-]$,
that gives a deformation of parafermionic algebra
of order $2p$.
At $p=1$ this algebra turns into the standard
parafermionic algebra of order $2$.

The (anti)commutation relations (\ref{6}), (\ref{7})
can be presented as 
$
\{f^+,f^-\}=F(N+1)+F(N),
$
$
[f^-,f^+]=F(N+1)-F(N),
$
with function $F(N)=N(-1)^N+(p+\frac{1}{2})(1-(-1)^N)$,
where $N=I_3+p$ is the number operator.
This means that deformed parafermionic algebra (\ref{8})
belongs to the class of generalized deformed parafermionic 
algebras introduced by Quesne \cite{que}.
Note also that the deformed $su(2)$ algebra
(\ref{7}) can be realized via the generators
of the standard $su(2)$ algebra, $[J_+,J_-]=2J_3$,
$[J_3,J_\pm]=\pm J_\pm$, taken in $(2p+1)$-dimensional
representation, $J_3^2+\frac{1}{2}\{J_+,J_-\}=p(p+1)$,
according to the prescription \cite{defsu,np1}:
$I_3=J_3$, $I_-=(I_+)^\dagger=
J_-\Phi(J_3)$ with
$\Phi(J_3)=[2p+1+(-1)^N(2J_3-1)]\cdot[2N(p-J_3+1)]^{-1}$.

The linear combinations of the 
initial creation-annihilation operators, 
$
{\cal L}_1=\frac{1}{\sqrt{2}}(a^+~+~a^-)$ and
$
{\cal L}_2=\frac{i}{\sqrt{2}}(a^+-a^-)
$,
satisfy the commutation relations
$
[{\cal L}_\alpha,{\cal L}_\beta]=i\epsilon_{\alpha\beta}
(1+\nu R)
$, 
$
\epsilon_{\alpha\beta}=-\epsilon_{\beta\alpha},
$
$\epsilon_{12}=1$,
and 
are hermitian in the case of infinite-dimensional representations
and self-conjugate with respect to the scalar product (\ref{5})
in the case of finite dimensional representations of RDHA.
These operators
together with quadratic operators $J_\mu$, $\mu=0,1,2,$
$
J_0=\frac{1}{4}\{a^+,a^-\},
$
$
J_1\pm i J_2=J_\pm=\frac{1}{2}(a^\pm)^2,
$
form the set of
generators of $osp(1|2)$ superalgebra:
$
\{{\cal L}_\alpha,{\cal L}_\beta\}
=4i(J\gamma)_{\alpha\beta},
$
$
[J_\mu,J_\nu]=-i\epsilon_{\mu\nu\lambda}J^\lambda$,
$
[J_\mu,{\cal L}_\alpha]=
\frac{1}{2}(\gamma_\mu)_\alpha{}^{\beta}{\cal L}_\beta.
$
Here $\gamma$-matrices appear in the Majorana representation,
$
(\gamma^{0})_{\alpha}{}^{\beta}=-(\sigma^{2})_{\alpha}{}^{\beta},
$
$
(\gamma^{1})_{\alpha}{}^{\beta}=i(\sigma^{1})_{\alpha}{}^{\beta},
$
$
(\gamma^{2})_{\alpha}{}^{\beta}=i(\sigma^{3})_{\alpha}{}^{\beta}.
$
Hence, 
$J_\mu$ are even and ${\cal L}_\alpha$ are odd generators 
of the superalgebra with $J_\mu$ forming 
$so(2,1)\sim sl(2,R)$ subalgebra and
${\cal L}_\alpha$ being an $so(2,1)$ spinor.
The $osp(1|2)$ Casimir operator 
$
{\cal C}\equiv J^\mu J_\mu-\frac{i}{8}{\cal L}^\alpha {\cal L}_\alpha,
$
$J^\mu=\eta^{\mu\nu}J_\nu,$ $\eta^{\mu\nu}=diag(-,+,+)$,
${\cal L}^\alpha=\epsilon^{\alpha\beta}{\cal L}_\beta$,
$\epsilon_{\alpha\gamma}\epsilon^{\gamma\beta}=-\delta_\alpha^\beta$,
takes the fixed value ${\cal C}=\frac{1}{16}(1-\nu^2)$.
Therefore,  every infinite- or finite-dimensional representation
of RDHA supplies us with the corresponding irreducible
representation of $osp(1|2)$ superalgebra revealing
its universality in some another but related aspect \cite{np1}.
Every such  representation is reducible
with respect to the action of the $so(2,1)$ generators $J_\mu$:
$
J^2=J^\mu J_\mu=-\hat{\alpha}(\hat{\alpha}-1),
$
where
$
\hat{\alpha}=\frac{1}{4}(1+\nu R).
$
Hence, $J_\mu$ act irreducibly on even, `$+$', and odd,
`$-$',  subspaces 
spanned by the states $|k\rangle_+$ and $|k\rangle_-$,
$
J^2|k>_\pm=-\alpha_\pm(\alpha_\pm-1)|k>_\pm,
$
where 
$\alpha_+=\frac{1}{4}(1+\nu)$,
$\alpha_-=\alpha_++1/2$
and 
$
J_0|k>_\pm=(\alpha_\pm+k)|k>_\pm,$ 
$
k=0,1,\ldots.
$
For infinite-dimensional representations of RDHA
($\nu>-1$), this  gives the direct sum of infinite-dimensional
unitary irreducible representations of $sl(2,R)$, ${\cal
D}^+_{\alpha_+}\oplus{\cal D}^+_{\alpha_-}$, being
half-bounded 
representations of the  discrete series characterized by parameters
$\alpha_+>0$ and $\alpha_->1/2$ \cite{sl2}.
In the case of finite-dimensional representations 
of RDHA one gets
$
J^2|l>_\pm=-j_\pm(j_\pm+1)|l>_\pm,
$
$j_+=p/2$, $j_-=(p-1)/2$,
where $l=0,1,\ldots,p$ for
$|l>_+$ 
and $l=0,1\ldots,p-1$ for
$|l>_-$.
Thus, we have the direct sum of
spin-$j_+$ and spin-$j_-$ finite-dimensional representations
with $so(2,1)$ spin parameter shifted in $1/2$,
where the operator $J_0$ has the spectra
$j_0=(-j_+,-j_++1,\ldots,j_+)$ and $j_0=(-j_-,-j_-+1,\ldots,
j_-)$ \cite{mpla2}.

The described $osp(1|2)$ superalgebraic construction 
can be extended to the OSp(2$|$2) supersymmetry.
Indeed, defining the operators
$\Delta=-\frac{1}{2}(R+\nu)$ and 
$Q^+=a^+\Pi_-$, 
$Q^-=a^-\Pi_+$,
$S^+=a^+\Pi_+$,
$S^-=a^-\Pi_-$,
we find that operators $J_\mu$ and $\Delta$
are even generators of $osp(2|2)$ superalgebra,
forming its $sl(2,R)\times u(1)$ subalgebra,
whereas $Q^\pm$ and $S^\pm$
are its odd generators \cite{ann,mpla2}.
In terms of the latter, $osp(1|2)$ odd generators
are presented as ${\cal L}_1=\frac{1}{\sqrt{2}}
(Q^++Q^-+S^++S^-)$,
${\cal L}_2=\frac{i}{\sqrt{2}}(Q^+-Q^-+S^+-S^-)$. 
This more broad OSp(2$|$2) supersymmetry was
revealed in ref. \cite{ann} as a dynamical symmetry
of bosonized supersymmetric 
quantum mechanical 2-body Calogero model.
Let us also note that the pair of odd generators
$Q^+$ and $Q^-$ together with even generator 
$H_+=2J_0+\Delta$ form $s(2)$ superalgebra,
$Q^{\pm2}=0$, $\{Q^+,Q^-\}=H_+$,
$[Q^\pm,H_+]=0$, 
whereas $S^+$ and $S^-$ are odd generators
of $s(2)$ superalgebra with even generator $H_-=2J_0-\Delta$.
This latter observation \cite{mac1,prep} was a starting point for 
bosonization of Witten supersymmetric quantum mechanics \cite{wit}
realized in ref. \cite{mpla1}.

To conclude the discussion of the formal algebraic
aspects, let us note that RDHA can be presented
in the form related to the generalized statistics \cite{guon}.
To this end, let us define the operators
$c^-=a^-G^{-1/2}_\nu(R)$,
$c^+=G^{-1/2}_\nu(R)a^+$,
where $G_\nu(R)=|1-\nu R|$, $\nu\neq 1$.
They give the normalized form
of the RDHA \cite{mpla2},
$
c^-c^+-g_\nu c^+ c^-=1,
$
with $g_\nu=(1-\nu)^R (1+\nu)^{-R}$ in the case
$-1< \nu<1$,
or 
$
c^- c^+-g_\nu c^+ c^-=R
$
with $g_\nu=(\nu-1)^R(1+\nu)^{-R}$ for $\nu>1$
and $g_\nu=p^R(1+p)^{-R}$
for $\nu=-(2p+1)$ (cp. with guon commutation relations
\cite{guon}).
The limit $|\nu|\rightarrow\infty$ in both cases 
$\nu>1$ and $\nu=-(2p+1)$ leads to the algebra
\begin{equation}
c^-c^+-c^+c^-=R,\quad 
\{R,c^\pm\}=0,\quad
R^2=1.
\label{9}
\end{equation}
This algebra has two-dimensional irreducible 
representation with reflection operator 
realized as $R=1-2c^+c^-$ that reduces eqs. (\ref{9})
to the standard fermionic anticommutation relations,
$\{c^+,c^-\}=1$, $c^{\pm2}=0$.
Therefore, the fermionic algebra can be obtained
as a limit case of RDHA. 
The commutation relations  (\ref{9}) can be generalized
to the algebra with phase operator \cite{mpla2},
\begin{equation}
[a,\bar{a}]={\cal R},\quad
{\cal R}^p=1,\quad {\cal R} a=qa{\cal R},\quad
{\cal R}\bar{a}=q^{-1}\bar{a}{\cal R},
\label{10}
\end{equation}
where $q=e^{-i2\pi/p}$, $p=2,3,\ldots$.
Via the transformation
$c=q^{-1/2}a{\cal R}^{-1/2}$,
$\bar{c}=q^{-1/2}{\cal R}^{-1/2}\bar{a}$,
the algebra (\ref{10}) can be related to the $q$-deformed Heisenberg
algebra $c\bar{c}-q\bar{c}c=1$ \cite{qdef}
with deformation parameter $q$ being the primitive root
of unity.

\section{Universal spinor set of field equations and 3D SUSY}
Now we turn to some applications
of the RDHA algebra \cite{mpla2,mpla?}. Namely, we shall employ
the described infinite- and finite-dimensional
representations for the construction of the universal
minimal set of linear differential equations
describing either ordinary integer and half-integer
$(2+1)$-dimensional spin-$j$ fields or anyons.
We shall also generalize the construction 
for the case of some 3D SUSY field systems and find
the corresponding field actions.

So, let us introduce the field $\Psi=\Psi^n(x)$ 
carrying some irreducible representation of RDHA.
The corresponding total angular momentum operator has
the form 
$M_\mu=-\epsilon_{\mu\nu\lambda}x^\nu P^\lambda
+ J_\mu$, $P_\mu=-i\partial_\mu$, and
generates Lorentz transformations,
$\Psi(x)\rightarrow \Psi'(x')=\exp(iM^\mu\omega_\mu)\Psi(x)$,
which are specified by parameters $\omega_\mu$.
In the chosen fixed representation of RDHA we introduce
the self-conjugate spinor operator 
\begin{equation}
{\cal Q}_\alpha={\cal Q}^\dagger_\alpha=
R{\cal P}_\alpha +\epsilon m{\cal L}_\alpha
\label{11}
\end{equation}
with $\epsilon=\pm$,
$
{\cal P}_\alpha=
(P\gamma)_\alpha{}^\beta {\cal L}_\beta
$
and $m$ a mass parameter.
Now we postulate the covariant set of field equations, 
\begin{equation}
{\cal Q}_\alpha\Psi=0.
\label{12}
\end{equation}
To clarify the physical content of the field $\Psi$ subject to 
eq. (\ref{12}), we
decompose it into irreducible $sl(2,R)$ components
$\Psi_\pm$,
$\Psi=\Psi_++\Psi_-$, $R\Psi_\pm=\pm\Psi_\pm$.
As a consequence of the equations (\ref{12})
and the (anti)commutation relations
\[
\{{\cal Q}_\alpha,{\cal Q}_\beta\}=
4i(P^2+m^2)(J\gamma)_{\alpha\beta}
-8i(\Delta_+\Pi_+
+\Delta_-\Pi_-)
(P\gamma)_{\alpha\beta},
\]
\[
[{\cal Q}_\alpha,{\cal Q}_\beta]=-
({\cal {\cal Q}}^\rho {\cal {\cal Q}}_\rho)
\epsilon_{\alpha\beta},
\quad
{\cal Q}^\rho {\cal Q}_\rho=i(P^2+m^2)(1+\nu R)
+8i\epsilon m
(\Delta_+\Pi_+
-\Delta_-\Pi_-),
\]
$
\Delta_\pm=PJ-\epsilon m\frac{1}{4}(\nu\pm 1),
$
we find that $\Psi_-=0$, whereas `even' component
$\Psi_+$ satisfies Klein-Gordon,
$(P^2+m^2)\Psi_+=0$, and spin,
$(PJ-s_+m)\Psi=0$, equations,
where the spin value is defined by 
the deformation parameter corresponding to the chosen 
irreducible representation of RDHA, $s_+=\epsilon\frac{1}{4}(1+\nu)$.
Hence, the spinor set of equations (\ref{12})
describes ordinary integer, $s_+=-\epsilon k$, or half-integer,
$s_-=-\epsilon (k+1/2)$, spin fields under the choice of
$\nu=-(4k+1)$, $k=1,2,\ldots,$ and $\nu=-(4k+3)$, $k=0,1,\ldots,$
respectively.
On the other hand, in the case of infinite-dimensional
representations ($\nu>-1$), the basic equations
describe the fields with arbitrary value of spin
$s_+\in {\bf R}$,
$s_+\neq 0$. 
In this case the spin equation, $(PJ-s_+m)\Psi=0$,
is the (2+1)-dimensional analog of the Majorana equation
\cite{plnp,maj}, whereas the basic spinor set of equations 
(\ref{12}) is some analog of the 4D Dirac positive-energy 
set of linear differential equations  \cite{dir}.

Therefore, the $R$-deformed Heisenberg algebra
gives a possibility to construct a universal minimal spinor set of linear
differential equations (\ref{11}), (\ref{12}) 
describing spin-$j$ fields and anyons \cite{mpla?}.

As we have seen,
any irreducible representation of the  $osp(1|2)$
superalgebra
is the direct sum of two irreducible 
representations of $so(2,1)$ subalgebra.
The latter are
specified by the parameters $\alpha_+$ and $\alpha_-$, or
$j_+$ and $j_-$ related as $\alpha_- -\alpha_+=j_+-j_-=1/2$.
So, one can try to modify the set of
linear differential equations (\ref{11}), (\ref{12})
in such a way that it would have
nontrivial solutions not only in `$+$' subspace, but also in
`$-$' subspace. If these states will have equal mass but their
spin will be shifted for $\Delta s=1/2 ({\rm mod}\, n)$, we shall
have the states forming a supermultiplet. 
Such a modification can be realized, but only for two
special cases corresponding to $\nu=-5$ and $\nu=-7$.
The reason is that 
the consistent system
of Klein-Gordon and spin equations cannot be obtained 
independently in the odd subspace.
But for those two cases one can 
arrive at the linear spin equation only which itself
will generate the Klein-Gordon equation.
Such corresponding linear spin equation is the Dirac equation
or the equation for topologically massive vector 
U(1) gauge field \cite{u1}.
We present here the final result whereas the details 
can be found in ref. \cite{mpla?}.

The 
self-conjugate spinor linear differential operator ${\cal D}_\alpha=
{\cal D}^\dagger_\alpha$ generating the corresponding set
of equations 
\begin{equation}
{\cal D}_\alpha\Psi=0
\label{13}
\end{equation}
is given by
\begin{equation}
{\cal D}_\alpha=(p-\frac{1}{2})\epsilon m{\cal L}_\alpha-{\cal J}_\alpha
+\frac{1}{2}(1-pR){\cal P}_\alpha,
\label{14}
\end{equation}
where $p=2,3$ correspond to $\nu=-5,-7$.
In these two cases we have supermultiplets with
spin content ($s_+=\epsilon$, $s_-=\frac{1}{2}\epsilon$)
for $\nu=-5$ and 
($s_+=\frac{3}{2}\epsilon$,
$s_-=\epsilon$) for $\nu=-7$.
It is interesting to note that in both cases, $p=2,3,$
the spinor supercharge operator
${\cal Q}_\alpha$, $\{{\cal Q}_\alpha, {\cal D}_\beta\}\approx 0$,
which transforms the corresponding supermultiplet components, 
coincides with operator (\ref{11}) taken in 5- and 7-dimensional
representations.
The weak equality means
that the left hand side turns into zero on the 
physical subspace given by eqs. (\ref{13}),
and we have the following
typical superalgebraic relations: 
$
\{{\cal Q}_\alpha,{\cal Q}_\beta\}
\approx-4i\epsilon m
(p+2)(P\gamma)_{\alpha\beta},
$
$
[P_\mu,{\cal Q}_\alpha]=0.
$
If we not require that
${\cal D}_\alpha^\dagger={\cal D}_\alpha$,
we shall have a possibility to get supermultiplets with the
spin content
$(s_+=\epsilon,$ $s_-=-\frac{1}{2}\epsilon)$ 
and with the spin shift $|\Delta s|=|s_+-s_-|=3/2$ 
for $\nu=-5$,
and $(s_+=\frac{3}{2}\epsilon,$ $s_-=-\epsilon$) 
with $|\Delta s|=5/2$ for $\nu=-7$.
The corresponding operators,
having the property
${\cal D}^\dagger_\alpha={\cal D}_\alpha+{\cal P}_\alpha$,
are given by
\begin{equation}
{\cal D}_\alpha=-\frac{1}{2}\epsilon m{\cal L}_\alpha-{\cal J}_\alpha
-\frac{1}{2}R{\cal P}_\alpha,\quad \nu=-5;
\quad
{\cal D}_\alpha=-\frac{1}{2}
\epsilon m{\cal L}_\alpha-\frac{1}{2}{\cal J}_\alpha
-\frac{1}{4}(1+R){\cal P}_\alpha,\quad \nu=-7.
\label{14*}
\end{equation}

Since the universal equations (\ref{12}) as well as 
equations for 3D SUSY field systems (\ref{13})
are two equations for one multi- or infinite-component
field, the corresponding action 
$
{\cal A}=\int Ld^3x
$
has to contain some auxiliary fields.
The Lagrangian leading to the 
equations (\ref{12}) for basic field $\Psi$ can be chosen as \cite{mpla?}
\begin{equation}
L=({\bar \Psi}+{\bar \chi}{}^\alpha {\cal Q}_\alpha)
(\Psi+{\cal Q}_\beta\chi^\beta)-{\bar \Psi}\Psi,
\label{15}
\end{equation}
where 
${\bar \Psi}=\Psi^\dagger$,
${\bar \chi}{}^{\alpha}=\chi^{\dagger}{}^{\alpha}$
for $\nu>-1$, whereas
${\bar \Psi}=\Psi^\dagger\hat{\eta}$,
${\bar \chi}{}^{\alpha}=\chi^{\dagger}{}^{\alpha}
\hat{\eta}$ for $\nu=-(2p+1)$, that
guarantees the reality of the Lagrangian.
Besides the basic equations (\ref{12}),
the action leads to the equations 
${\cal Q}_\alpha\chi^\alpha=0$ and is invariant 
with respect to the transformations
$\chi^\alpha\rightarrow \chi^\alpha+\Pi^{\alpha\beta}\Lambda_\beta$,
where 
$
\Pi_{\alpha\beta}=({\cal Q}^\sigma {\cal Q}_\sigma)\epsilon_{\alpha\beta}
-{\cal Q}_\alpha {\cal Q}_\beta,
$
$
{\cal Q}^\alpha\Pi_{\alpha\beta}=0.
$
This means that fields $\chi^\alpha$ play the role 
of auxiliary fields having no independent dynamics.
The same equations can be obtained also 
from the action with the linear form of
Lagrangian,
\begin{equation}
L'={\bar \chi}{}^\alpha {\cal Q}_\alpha\Psi+
{\bar \Psi}{\cal Q}_\alpha\chi^\alpha.
\label{16}
\end{equation}
The change of ${\cal Q}_\alpha$ for ${\cal D}_\alpha$ in
Lagrangians (\ref{15}) and (\ref{16}) will give the action
functionals for the special SUSY cases (\ref{14}) 
corresponding to $\nu=-5$ and $\nu=-7$.
For SUSY systems given by eq. (\ref{14*}),
the corresponding quadratic and linear
Lagrangians are
$
L=({\bar \Psi}+{\bar \chi}{}^\alpha {\cal D}_\alpha)
(\Psi+{\cal D}^\dagger_\beta\chi^\beta)-{\bar \Psi}\Psi
$
and 
$
L'={\bar \chi}{}^\alpha {\cal D}_\alpha\Psi+
{\bar \Psi}{\cal D}^\dagger_\alpha\chi^\alpha.
$

The described first and second order field Lagrangians
can be used as a departing point for realization of the second
quantization of the corresponding (2+1)-dimensional 
field systems. This, in particular, could clarify the question on
the spin-statistics relation for the 
described fractional spin fields.


\begin{thebibliography}{**}

\bibitem{ok}
Y. Ohnuki and S. Kamefuchi,
{\it Quantum Field Theory and Parastatistics},
University Press of Tokyo, 1982.

\bibitem{np1}
M. S. Plyushchay, {\it Nucl. Phys.} {\bf B491} (1997) 619
[{\bf hep-th/9701091}].

\bibitem{wig}
E. P. Wigner, {\it Phys. Rev.} {\bf 77} (1950) 711.

\bibitem{mac2}
A. J. Macfarlane, {\it Generalized Oscillator Systems and Their 
Parabosonic Interpretation}, in:
Proc. Inter. Workshop on Symmetry Methods in Physics,
eds. A. N. Sissakian, G. S. Pogosyan and 
S. I. Vinitsky (JINR, Dubna, 1994) p. 319.


\bibitem{green}
H. S. Green, {\it Phys. Rev.} {\bf 90} (1953) 270.

\bibitem{volkov}
D. V. Volkov, {\it Sov. Phys. JETP} {\bf 9} (1959) 1107;
{\bf 9} (1960) 375.

\bibitem{gg}
O. W. Greenberg, {\it Phys. Rev. Lett.} {\bf 13} (1964) 598;\\
O. W. Greenberg and A. M. L. Messiah, 
{\it Phys. Rev.} {\bf B138}  (1965) 1155;\\
A. B. Govorkov, {\it Theor. Math. Phys.} {\bf 54}
(1983) 234.

\bibitem{poly}
A. P. Polychronakos, {\it Phys. Rev. Lett.}
{\bf 69} (1992) 703.

\bibitem{calog}
L. Brink, T. H. Hansson, S. Konstein, and M. A. Vasiliev,
{\it Nucl. Phys.} {\bf B401} (1993) 591.

\bibitem{yang}
L. M. Yang, {\it Phys. Rev.} {\bf 84} (1951) 788. 


\bibitem{mac1}
T. Brzezinski, I. L. Egusquiza, and A. J. Macfarlane,
{\it Phys. Lett.} {\bf B311} (1993) 202.

\bibitem{prep}
M. S. Plyushchay, {\it Supersymmetry
without Fermions}, {\bf hep-th/9404081}.


\bibitem{mpla1}
M. S. Plyushchay, {\it Mod. Phys. Lett.}
{\bf A11} (1996) 397.


\bibitem{ann}
M. S. Plyushchay, {\it Ann. Phys.} {\bf 245} (1996) 339.

\bibitem{plb0}
M. S. Plyushchay, {\it Phys. Lett.} {\bf B320} (1994) 91.



\bibitem{plb1}
M. S. Plyushchay, {\it Phys. Lett.} {\bf B248} (1990) 107.

\bibitem{jn}
R. Jackiw and V. P. Nair, 
{\it Phys. Rev.} {\bf D43} (1991) 1933. 

\bibitem{plnp}
M. S. Plyushchay,
{\it Phys. Lett.} {\bf B262} (1991) 71;
{\it Nucl. Phys.} {\bf B362} (1991) 54.

\bibitem{gru}
M. S. Plyushchay,
{\it Phys. Lett.} {\bf B273} (1991) 250;
{\it Int. J. Mod. Phys.} {\bf A7} (1992) 7045.

\bibitem{volsor}
D. P. Sorokin and D. V. Volkov,
{\it Nucl. Phys.} {\bf B409} (1993) 547.



\bibitem{mpla2}
M. S. Plyushchay, {\it Mod. Phys. Lett.}
{\bf A11} (1996) 2953.


\bibitem{mpla?}
M. S. Plyushchay, {\it R-deformed Heisenberg algebra,
anyons and $d=2+1$ supersymmetry},
{\bf hep-th/9705034},
to appear in {\it Mod. Phys. Lett.} {\bf A12}.

\bibitem{fili}
A. T. Filippov, A. P. Isaev, and A. P. Kurdikov,
{\it Mod. Phys. Lett.} {\bf A7} (1992) 2129;
{\it Teor. Mat. Fiz.} {\bf 94} (1993) 213;\\
N. Fleury and M. Rausch de Traubenberg,
{\it J. Math. Phys.} {\bf 33} (1992) 3356;
{\it Adv. Appl. Clif. Alg.} {\bf 4} (1994) 123.

\bibitem{que}
C. Quesne, {\it Phys. Lett.} {\bf A193} (1994) 245.

\bibitem{defsu}
T. Curtright and C. Zachos, {\it Phys. Lett.} {\bf B243}
(1990) 237;\\
A. P. Polychronakos, {\it Mod. Phys. Lett.}
{\bf A5} (1990) 2325;\\
M. Ro\v{c}ek, {\it Phys. Lett.} {\bf B255} (1991) 554.


\bibitem{sl2}
M. S. Plyushchay,
{\it J. Math. Phys.} {\bf 34} (1993) 3954.

\bibitem{wit}
E. Witten,
{\it Nucl. Phys.} {\bf B188} (1981) 513;
{\bf B202} (1982) 253.

\bibitem{guon}
R. Scipioni, {\it Phys. Lett.} {\bf B327} (1994) 56;
{\it Nuovo Cim.} {\bf B109} (1994) 479;
L. De Falco, R. Mignani and R. Scipioni,
{\it Nuovo Cim.} {\bf A108} (1995) 1029.

\bibitem{qdef}
M. Arik and D. D. Coon,
{\it J. Math. Phys.} {\bf 17} (1976) 524;\\
V. Kuryshkin, {\it Ann. Fond. L. de Broglie} {\bf 5}
(1980) 111;\\
A. J. Macfarlane, {\it J. Phys. A} {\bf 22} (1989) 4581;\\
L. C. Biedenharn, {\it J. Phys. A} {\bf 22} (1989) L873.

\bibitem{maj}
E. Majorana, {\it Nuovo Cim.} {\bf 9} (1932) 335;\\
W. R\"uhl, {\it Comm. Math. Phys.} {\bf 6} (1967) 312.

\bibitem{dir}
P. A. M. Dirac, {\it Proc. Roy. Soc. London Ser. A}
{\bf 322} (1971) 435, {\bf 328} (1972) 1572. 


\bibitem{u1}
R. Jackiw and S. Templeton, {\it Phys. Rev.} {\bf D23} (1981) 2291;\\
J. Schonfeld, {\it Nucl. Phys.} {\bf B185} (1981) 157;\\
S. Deser, R. Jackiw and S. Templeton, 
{\it Phys. Rev. Lett.} {\bf 48} (1982) 975.


\end{thebibliography}
\end{document}